# Classifying Intermediate-redshift Galaxies in SDSS: Alternative Diagnostic Diagrams

Léa M. Feuillet[1], Marcio Meléndez[2], Steve Kraemer[1], Henrique R. Schmitt[3], Travis C. Fischer[4], and James N. Reeves[1]

[1] Institute for Astrophysics and Computational Sciences, Department of Physics, The Catholic University of America, Washington, DC 20064, USA; feuilletl@cua.edu  
[2] Space Telescope Science Institute, 3700 San Martin Drive, Baltimore, MD 21218, USA  
[3] Naval Research Laboratory, Remote Sensing Division, 4555 Overlook Avenue SW, Washington, DC 20375, USA  
[4] AURA for ESA, Space Telescope Science Institute, 3700 San Martin Drive, Baltimore, MD 21218, USA  
Received 2023 July 19; revised 2023 December 22; accepted 2023 December 23; published 2024 February 12

## Abstract

We select a sample of 1437 active galactic nuclei (AGN) from the catalog of the Sloan Digital Sky Survey galaxy properties from the Portsmouth group by detection of the high-ionization [Ne V] 3426 Å emission line. We compare the fluxes of [Ne III] 3869 Å, [O III] 5007 Å, [O II] 3726, 3728 Å, and [O I] 6300 Å to that of [Ne V]. All four lines show a strong linear correlation with [Ne V], although lines from ions with lower ionization potentials have a lower correlation coefficient. We investigate the use of two forbidden line ratio diagnostic diagrams that do not rely on H$\alpha$ in order to classify high-redshift galaxies. These use the [Ne III]/[O II] line ratio plotted against [O III]/[O I] and [O III]/[O II], respectively. We use photoionization modeling to characterize the behavior of the narrow-line region in AGN and star-forming regions and test the validity of our diagnostic diagrams. We also use a luminosity cutoff of log $L_{\rm [O\,III]}$ [erg s$^{-1}$] = 42, which lowers the contamination of the AGN region by star-forming galaxies down to 10% but does not remove green pea and purple grape galaxies from the AGN region. We also investigate the OHNO diagram, which uses [Ne III]/[O II] plotted against [O III]/H$\beta$. Using our new diagnostic diagrams, we are able to reliably classify AGN up to a redshift of $z \leqslant 1.06$ and add more than 822 new AGN to the [Ne V]-selected AGN sample.

*Unified Astronomy Thesaurus concepts:* Active galactic nuclei (16); Classification systems (253); Starburst galaxies (1570)

*Supporting material:* machine-readable table

## 1. Introduction

The ability to distinguish between active galactic nuclei (AGN) and star-forming galaxies (SFGs) is essential to obtaining a pure sample of AGN. Multiple kinds of studies rely upon knowing galaxy populations, such as galaxy evolution or the ionization of the intergalactic medium. Diagnostic diagrams use ratios of emission lines to distinguish AGN from SFGs in large samples. To review, active galaxies can be divided into multiple categories, including quasars, the most optically luminous AGN, and Seyfert galaxies, which are the most luminous local AGN. Seyfert galaxies may be further classified into types 1 and 2, which differ due to the viewing angle of the galaxy. A Seyfert 1 galaxy corresponds to an AGN whose broad-line region (BLR) is visible, leading to the spectra having both narrow and broad components to the emission lines. In type 2 AGN, the BLR is obscured by the torus, and only the narrow-line region (NLR) can be observed (Khachikian & Weedman 1974). Another type of AGN is low-ionization nuclear emission-line regions (LINERs; Heckman 1980), which have strong narrow emission lines like AGN but are characterized by lower ionization. Finally, composite galaxies show both strong star formation and weak AGN features within the same galaxy.

The most famous and widely used diagnostic diagrams are the Baldwin, Phillips & Terlevich (BPT) diagram (Baldwin et al. 1981) and the Veilleux and Osterbrock (VO) diagrams (Veilleux & Osterbrock 1987). The [N II]-based BPT diagram uses four emission lines, plotting [O III] 5007 Å/H$\beta$ against [N II]/H$\alpha$. On such a plot, one would find the Seyfert galaxies in the top right corner, having both high [O III]/H$\beta$ and high [N II]/H$\alpha$. The LINERs, on the other hand, would have a comparatively lower [O III]/H$\beta$ ratio and thus would be found on the lower right of the plot. Additionally, the SFGs are present on the left side, with a wide range of [O III]/H$\beta$ values but low [N II]/H$\alpha$. Finally, composite galaxies are found close to the delimitation line between the AGN and the SFGs. This method was used by the Portsmouth group to assign BPT types to all galaxies, with a BPT type of "BLANK" given for those missing one of the four necessary emission lines (Thomas et al. 2013).

BPT and VO diagrams do come with limitations, especially when it comes to redshift constraints. As these diagrams rely on the H$\alpha$ line, they may only be used in the Sloan Digital Sky Survey (SDSS) and most optical surveys up to a redshift of 0.57. Multiple diagnostic diagrams have been created since then in order to solve this very issue (Stasińska et al. 2006; Lamareille 2010; Juneau et al. 2011, 2014; Trouille et al. 2011; Yan et al. 2011). As instruments become more powerful and are able to capture objects that are further away, introducing diagnostic diagrams able to classify galaxies at higher redshift is becoming increasingly relevant.

The use of the optical [Ne V] 3426 Å (hereafter [Ne V]) emission line has not been widely adopted for AGN classification, unlike the IR [Ne V] lines (e.g., Abel & Satyapal 2008; Satyapal et al. 2008; Goulding & Alexander 2009). Ne$^{+4}$ has a high ionization potential (IP; 97 eV) and can only be produced by high-energy phenomena such as shocks and the presence of AGN. Even the hottest main-sequence stars do not emit photons energetic enough to produce [Ne v]. This has been shown by Baldwin et al. (1981), as none of the H II regions investigated showed any amount of [Ne v] within their spectra. The use of [Ne v] for AGN selection is also demonstrated in, e.g., Schmidt et al. (1998), Gilli et al. (2010), and Mignoli et al. (2013).

This paper is outlined as follows. Section 2 details the selection process used to obtain the samples adopted throughout the paper. Section 3 demonstrates different ways of classifying galaxies using a total of three different alternative diagnostic diagrams. Specifically, we demonstrate the usefulness of the emission lines used in Section 3.1 and consider the physical reasons behind the capabilities of our diagnostic diagrams in Section 3.2. We introduce a third axis in the form of a luminosity cutoff in Section 3.3 and discuss our diagrams' limitations in Section 3.4. We also show the possible applications of using the diagrams in Section 4. We use Wilkinson Microwave Anisotropy Probe 7 yr results cosmology throughout, with $H_0 = 71.0$ km s$^{-1}$ Mpc$^{-1}$ and $\Omega_0 = 0.734$ (Komatsu et al. 2011).

## 2. Samples and Data Selection

We use SDSS Data Release 12 (Alam et al. 2015) galaxy data extracted and made available by the Portsmouth group[5] (Thomas et al. 2013). SDSS is an optical survey with an extensive galaxy catalog, which allows us to use the [Ne v] optical line to classify our AGN among a large initial sample. The emission line fluxes and other spectral information come from the Baryon Oscillation Spectroscopic Survey (BOSS). The optical fibers have a diameter of 2″, corresponding to physical diameters between 7 and 23 kpc for our redshift range. Therefore, the BOSS spectra used in the present paper not only encompass the central AGN but will also have some contribution from the host galaxy due to the relatively high redshift of the samples.

We use the emissionLinesPort (Thomas et al. 2013) table to get emission line fluxes and their uncertainties, as well as redshifts for the galaxies, which are all calculated using the Graphical Astrophysics code for N-body Dynamics And Lagrangian Fluids (Sarzi et al. 2006) and Penalized PiXel-Fitting code (Cappellari & Emsellem 2004), which are both publicly available.

We reduced the full sample of nearly 1.5 million galaxies into three subsamples based on the [N II]-based BPT classifications provided by the Portsmouth group: the star-forming sample; the AGN sample containing the galaxies classified as Seyfert,[6] composite, or LINER; and the unclassified sample. Starting at a redshift of $z = 0.57$, the H$\alpha$ line is shifted out of bounds of the SDSS spectrum range. As these galaxies cannot be plotted on a regular BPT diagram, they cannot be classified using this standard method and are therefore assigned a BPT type of "BLANK"[7] by the Portsmouth group (as shown in Figure 1). Finally, we have the [Ne v] sample, which is the result of imposing a signal-to-noise ratio (S/N) >3 for the [Ne v] emission line flux for our AGN sample. As the [Ne v] line is found toward the shorter-wavelength end of the BOSS spectra, we set a lower limit to the redshift of 0.168 to guarantee the presence of the [Ne v] line above a threshold of 4000 Å.

---
[5] https://www.sdss.org/dr12/spectro/galaxy_portsmouth/
[6] The Portsmouth group does not distinguish between these two types within their data set, fitting a single Gaussian to the emission lines without attempting to fit any potential broad components.
[7] Other galaxies that lack emission lines are also classified as "BLANK" regardless of redshift. However, our S/N cuts for various emission lines remove these cases from the unclassified sample, leaving only those with $z > 0.57$.

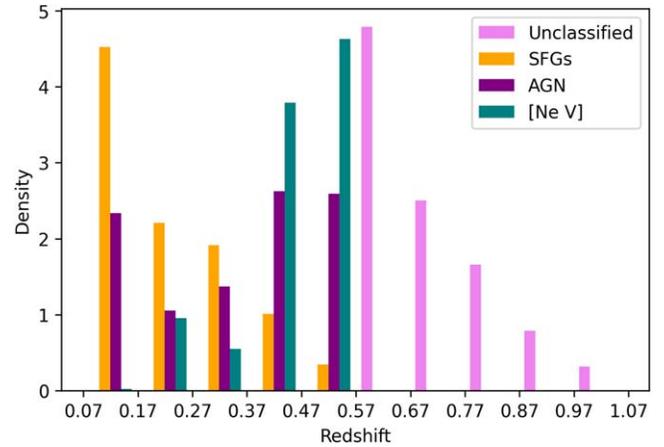

**Figure 1.** Redshift distribution for the unclassified, star-forming, AGN, and [Ne v] samples. The histograms are an approximation of the probability density function for each sample's distribution. The redshift values were calculated by the Portsmouth group using the models found in Maraston et al. (2006). We binned the redshift values in bins of width 0.1 between $z = 0.07$ and $z = 1.07$.

The emission line fluxes were corrected for interstellar reddening by using the observed and theoretical Balmer decrements (H$\alpha$/H$\beta$) and the Cardelli et al. (1989) reddening law for $R_v = 3.1$ to obtain the $E(B - V)$ values. The intrinsic H$\alpha$/H$\beta$ ratio in AGN can range from 2.9 (Osterbrock & Ferland 2006) to ≈3.1 for broad-line H$\alpha$/H$\beta$ (Dong et al. 2008). As the Seyfert galaxies are not distinguished between type 1s and type 2s in the Portsmouth sample, we use an average value of H$\alpha$/H$\beta$ = 3. For the unclassified sample, we instead use the H$\beta$/H$\gamma$ decrement with a theoretical value of 2.13, as H$\alpha$ is shifted out of bounds for these galaxies.

We additionally require an S/N greater than 3 in all four subsamples for the [Ne III] 3869 Å, [O III] 5007 Å, and [O II] 3726, 3728 Å emission lines (hereafter [Ne III], [O III], and [O II]). We also require S/N > 2 for the H$\beta$ emission line in all samples but the unclassified galaxies, as the error values are all set to 0, as well as an S/N > 2 for H$\gamma$ for the unclassified galaxies only. We set an upper limit of 0.57 to the redshift of all samples but the unclassified galaxies to ensure adequate wavelength coverage of H$\alpha$. When using [O I] 6300 Å (hereafter [O I]), we limit our sample to galaxies with S/N > 2 due to the relative weakness of the line. We do not remove galaxies from the original samples based on this requirement but create a subsample to use in that particular instance. An overview of the redshift distribution for the four subsamples is provided in Figure 1, and a summary of the samples and selection criteria is provided in Table 1. The AGN and SFGs are also plotted on the [N II]–BPT diagram in Figure 2, along with the Kewley et al. (2001), Kauffmann et al. (2003), and Schawinski et al. (2007) demarcation lines.

## 3. Alternative Diagnostic Diagrams

### 3.1. Flux Relationships

Diagnostic diagrams work on the basis that emission lines track the physical conditions within galaxies, which differ between SFGs and AGN. The differences in physical conditions are mainly due to the presence of the AGN, which provides a harder ionizing spectrum to ionize the gas in the inner regions of the host. In

**Table 1**
Summary of the Samples with the Sample Size and Selection Criteria Specified

| Sample | Sample Size | Selection Criteria |
| --- | --- | --- |
| Star-forming | 3911 | BPT type is star-forming |
| AGN | 2847 | BPT type is Seyfert, composite, or LINER |
| [Ne V] | 1437 | Included in AGN sample, [Ne V] S/N > 3, z > 0.168 |
| Unclassified | 1246 | BPT type is BLANK |

**Note.** We also require an S/N greater than 3 in all four subsamples for the [Ne III], [O III], and [O II] emission lines.

AGN, X-ray photons penetrate regions beyond the $H^+/H^0$ transition zone (Osterbrock & Ferland 2006). The resulting ionization, most importantly of $H^0$, produces suprathermal photoelectrons, creating an extended warm zone beyond the transition layer. Collisional excitation of ions and neutral atoms in the warm zone produces forbidden lines such as [O I] 6300, 6364 Å. Also, the collisional excitation of low-lying levels of $N^0$ creates conditions for photoionization by photons with energies $h\nu < 13.6$ eV, producing $N^+$ in the warm zone (e.g., Henry 1970). This results in enhancement of [N II] 6548, 6583 (see discussion in Kraemer & Harrington 1986). Overall, these effects increase the [O I]/Hα and [N II]/Hα ratios, subsequently separating AGN and star-forming regions in BPT diagrams. On the other hand, SFGs' emission lines originate from the H II regions of the galaxy.

We now look at the relationship between the line fluxes of [Ne III], [O III], [O II], and [O I] and that of [Ne V] in order of decreasing IP. These flux relationships inform us about the conditions in which the different lines form based on their correlation with [Ne V], which is a pure tracer of AGN activity.

For each flux relationship, we find the Pearson correlation coefficient (r) and the standard deviation (σ) for the regression line.[8] The resulting values are summarized in Table 2, along with additional physical information about each line. Both the correlation coefficient and the standard deviation find that the relationship is strongest between [Ne V] and [O III], closely followed by [Ne III], then [O II], and finally [O I] with the weakest correlation.

The [Ne V], [Ne III], and [O III] are all created in the same AGN $H^+$ zone, which explains the high correlation and low scatter between these lines. Although [O II] is also created in the same $H^+$ zone as the other lines (Osterbrock & Ferland 2006), it is also a known tracer of star formation, produced in the H II regions of the host galaxy (Kewley et al. 2004; Zhuang & Ho 2019). This significant contribution from the star-forming regions could be the source of the scatter in the relationship. The [O I] line has a significantly lower correlation coefficient compared to the other three lines, as producing strong [O I] requires X-rays in a partially ionized transition zone (Osterbrock & Ferland 2006). This line is therefore not as strongly correlated to [Ne V], as they are produced in different zones, independently of one another (Kraemer & Crenshaw 2000). Although they differ in the strength of their correlation, all four lines show a clear linear relationship with [Ne V], suggesting their usefulness in constructing alternative diagnostic diagrams.

---

[8] The standard deviation for the regression line is calculated using the following equation: $S_{y/x} = \sqrt{S_r/(n-2)}$, where $S_r = \sum_{i=1}^{n}(y_i - a_0 - a_1 x_i)^2$ is the sum of the squares of the residuals with respect to the regression line, $a_0$ and $a_1$ are the intercept and slope of the regression line, and $n$ is the number of data points.

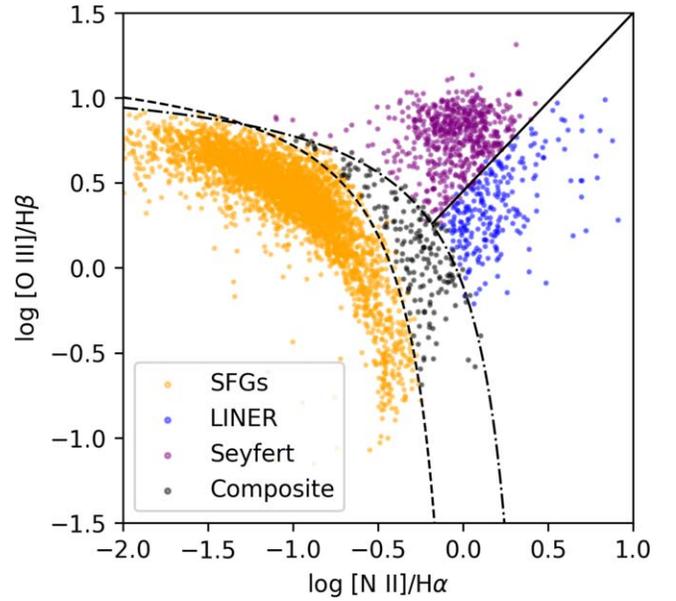

**Figure 2.** [N II]-based BPT diagram with our star-forming and full AGN samples. The dashed–dotted line corresponds to the Kewley et al. (2001) extreme starburst demarcation line, and the dashed line is the Kauffmann et al. (2003) line. The solid line is the Seyfert/LINER distinction line from Schawinski et al. (2007). The SFGs are in orange, the LINERs in blue, the Seyfert galaxies in purple, and the composite galaxies in black. The SFGs are concentrated toward higher [O III]/Hβ ratios, indicating the presence of higher-ionization H II regions.

### 3.2. The Forbidden Line Ratio Diagrams

Although the BPT diagram is the most commonly used diagnostic, it relies on the presence of Hα. In order to classify galaxies in which the Hα line is missing, due to either faulty flux determinations or the unavailability of the line due to a high redshift, other emission line diagnostics are needed.

Here, we first suggest a forbidden line ratio (FLR) diagnostic that can be used to distinguish between SFGs and AGN and utilizes the [Ne III]/[O II] versus [O III]/[O I] flux ratios. These particular ratios have not been used in conjunction in the past as a way to distinguish luminous AGN from SFGs. However, the [O I]/[O III] ratio was used by Heckman (1980) and, later on, Filippenko & Terlevich (1992) to classify LINERs. When plotting our star-forming and [Ne V] samples on the [O I]–FLR diagram, the two separate quite evidently (see Figure 3, left panel). The [Ne V] sample occupies the upper left section of the plot, and the star-forming one fills in the lower center and upper right sections. The demarcation line is chosen to maximize the percentage of [Ne V] galaxies and minimize the number of star-forming ones in the designated AGN region. The final equation for the demarcation line is

$$\log_{10}\left(\frac{[\text{Ne III}]}{[\text{O II}]}\right) > 1.25\left(\log_{10}\left(\frac{[\text{O III}]}{[\text{O I}]}\right) - 1.15\right)^2 - 0.95, \quad (1)$$

which includes 87% and 2% of the [Ne V] and star-forming samples, respectively, in the area above the parabola, which we refer to as the AGN region.

In order to model the behavior of the NLR in AGN and star-forming regions, we use the photoionization modeling code Cloudy (v. 17.0; Ferland et al. 2017) to constrain the physical conditions that produce the [Ne III]/[O II] and [O III]/[O I] ratios.

Table 2
Summary of Partial Correlation Coefficients and Standard Deviation around the Regression Line for the Four Emission Lines Analyzed with Respect to [Ne v]

| Line | IP (eV) | Transition | $n_c$ (cm$^{-3}$) | Slope | Intercept | $r$ | $\sigma$ |
|---|---|---|---|---|---|---|---|
| [Ne v] | 97 | $^3P_2$ - $^1D_2$ | 1.3e7[a] | ... | ... | ... | ... |
| [Ne III] | 41.0 | $^3P_2$ - $^1D_2$ | 9.5e6[a] | 0.96 | 0.15 | 0.91 | 0.24 |
| [O III] | 35.1 | $^3P_2$ - $^1D_2$ | 6.8e5[a] | 0.75 | 1.50 | 0.92 | 0.24 |
| [O II] | 13.6 | $^4S^0_{3/2}$ - $^2D^0_{3/2}$ | 1.5e4[a] | 0.89 | 0.83 | 0.89 | 0.34 |
| [O I] | NA | $^3P_2$ - $^1D_2$ | 5.0e5[b] | 0.55 | 0.58 | 0.56 | 0.60 |

**Notes.** IP is the ionization potential of the ion. $n_c$ is the critical density of the line, corresponding to the specified transition.
[a] Values obtained from Table 3.15 in Osterbrock & Ferland (2006).
[b] Value taken from Goldsmith (2019). The slope and intercept values correspond to the slope and intercept for the line of best fit of each distribution. $r$ is the Spearman rank for the correlation, and $\sigma$ is the standard deviation for the regression line.

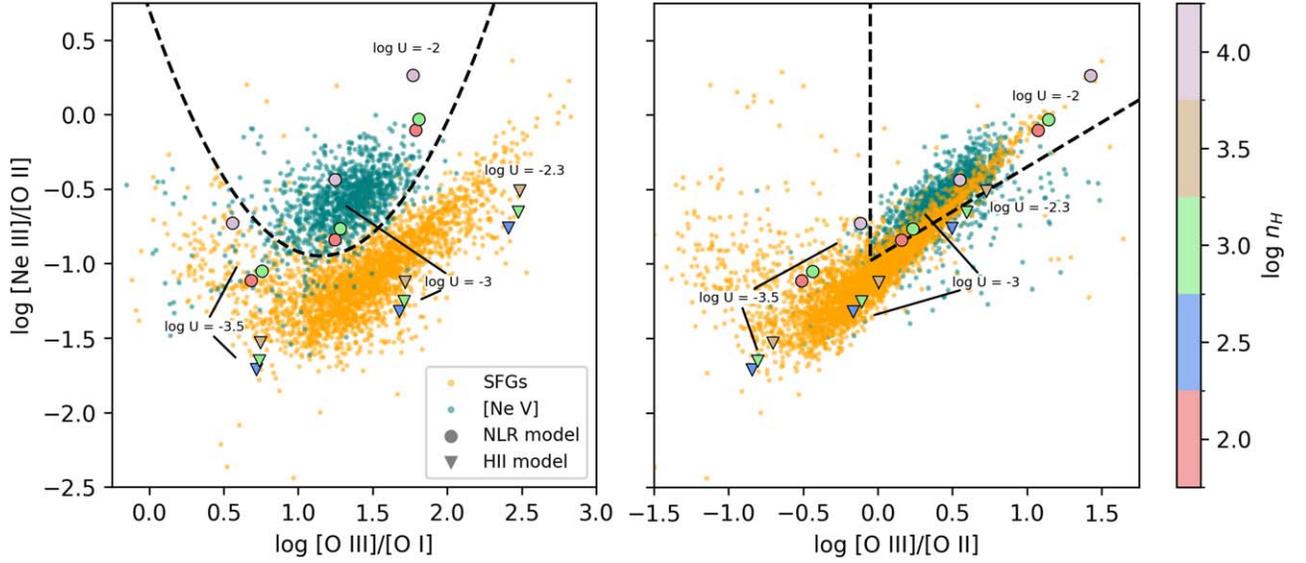

**Figure 3.** The two FLR diagrams considered here: [Ne III]/[O II] vs. [O III]/[O I] and [Ne III]/[O II] vs. [O III]/[O II]. The shape of the markers distinguishes between the NLR (circles) and H II region (triangles) models used to model the [Ne v] and star-forming samples, respectively. The ionization parameter used for a particular set is indicated in the figure, and the hydrogen density (log $n_H$) is indicated by the color of the marker. Left panel: [Ne III]/[O II] vs. [O III]/[O I] FLR ([O I]–FLR) diagram with the star-forming and [Ne v] samples and a demarcation line given by Equation (1) separating them. Right panel: [Ne III]/[O II] vs. [O III]/[O II] FLR ([O II]–FLR) diagram showing the star-forming and [Ne v] samples with the demarcation lines defined in Equations (3) and (4) as dashed black lines. The AGN are then found preferentially in the upper right section of the diagram.

We adopt an NLR model and assume the spectral energy distribution as described in Trindade Falcão et al. (2021). We use solar abundances, with exact values of C = −3.54, N = −4.17, O = −3.31, Ne = −3.94, Na = −5.78, Mg = −4.45, Al = −5.57, Si = −4.49, S = −4.88, Ar = −5.7, Ca = −5.7, Fe = −4.54, and Ni = −5.8, relative to hydrogen on a logarithmic scale. We define the ionization parameter as

$$U = \frac{Q}{4\pi c r^2 n_H}, \quad (2)$$

where $Q$ is the number of ionizing photons per second emitted by the central object, $r$ is the distance between the central object and the cloud, $c$ is the speed of light, and $n_H$ is the hydrogen number density (Osterbrock & Ferland 2006). Ferland & Netzer (1983) suggest that the ideal ionization parameter values range from $\approx 10^{-2}$ for Seyfert 1 and quasar NLRs to $\approx 10^{-2.5}$ for Seyfert 2 and $\approx 10^{-3.5}$ for LINERs. We thus vary the ionization parameter between log $U = -3.5$ and $-2$ and the hydrogen density within the range log $n_H = 2$–4 cm$^{-3}$. The model is radiation-bounded, and the computation stops when a temperature of 4000 K is reached. The results from this model are shown in Figure 3.

We then use Starburst99 models[9] and Cloudy to predict the [Ne III]/[O II] and [O III]/[O I] ratios in SFGs (Leitherer et al. 1999). Our Starburst99 simulation uses a continuous star formation rate (SFR) of 2.0 $M_\odot$ yr$^{-1}$, similar to the mean SFR of our star-forming sample[10] of 1.71 $M_\odot$ yr$^{-1}$, and a Salpeter initial mass function power law. Any other parameter is left to the default values prefilled in Starburst99, including the metallicity, which is kept at $Z = 0.014$ (solar). The Cloudy model assumes an age of 1 Myr and a closed and spherical geometry with a radius of log $R = 19$, the temperature is solved for, and the density is varied between log $n_H = 2.5$ and 3.5 cm$^{-3}$ in increments of 0.5 dex. We vary the ionization parameter between log $U = -3.5$ and $-2.3$. The upper limit to log $U$ for the H II region stems from the wind pressure dominating gravity as $U$ increases, which creates winds that push the gas out from the H II region (Yeh & Matzner 2012).

---
[9] https://www.stsci.edu/science/starburst99/docs/parameters.html
[10] We obtained the SFR data from the Portsmouth stellarMassStar-formingPort table.

Overall, the results for both the NLR and H II region reflect the behavior of our sample. The two tracks deviate significantly, and the NLR models are constrained within the AGN region, while the H II ones lie far below the demarcation line. However, as previously mentioned, we want a diagnostic diagram that only relies on emission lines with shorter wavelengths in order to increase the range in redshift we may use it for. Whereas [O I] does have a shorter wavelength than Hα, 92% of the unclassified sample also has the [O I] line shifted out of bounds.

Therefore, we now suggest a slightly different FLR diagram, using instead the [Ne III]/[O II] versus [O III]/[O II] flux ratios ([O II]–FLR), which have been investigated in the past (e.g., Pérez-Montero et al. 2007; Trouille et al. 2011; Levesque & Richardson 2014; Witstok et al. 2021). Levesque & Richardson (2014) used Starburst99 and Mappings III in the context of using [Ne III]/[O II] as an ionization parameter diagnostic in SFGs. [Ne III]/[O II] is also used in the Trouille, Barger, and Tremonti diagnostic diagram, where it is plotted against rest-frame g–z colors (Trouille et al. 2011). In Witstok et al. (2021), the authors plot [Ne III]/[O II] versus [O III]/[O II] flux ratios for both SFGs and Seyfert galaxies from SDSS DR7. They do so to investigate the tight relationship between the two ratios and consequently use [Ne III]/[O II] instead of the more commonly used [O III]/[O II] as an ionization parameter diagnostic for high-redshift galaxies. It is clear that the star-forming and Seyfert galaxy samples separate quite significantly in their plot, giving these ratios the potential to work as an AGN diagnostic diagram.

As SFGs have a higher relative [O II] emission, we expect them to be located toward the lower left of the diagram (see Figure 3, right panel). We have reddening-corrected all of our fluxes, as discussed in Section 2, as the [O III]/[O II] ratio is potentially more affected by reddening than [Ne III]/[O II]. We once again chose values for the demarcation lines that both maximized the number of [Ne V] AGN and minimized the number of SFGs contained in the AGN region. We found the optimized limits of

$$\log_{10}\left(\frac{[\text{O III}]}{[\text{O II}]}\right) > -0.05, \quad (3)$$

$$\log_{10}\left(\frac{[\text{Ne III}]}{[\text{O II}]}\right) > 0.60 \log_{10}\left(\frac{[\text{O III}]}{[\text{O II}]}\right) - 0.95. \quad (4)$$

Using these demarcation lines, the AGN can be found in the upper right section of the plot, which contains in total 80% of the [Ne V] sample and 19% of all SFGs.

Using the same parameters as previously discussed for the Cloudy models on our [O I]–FLR diagram, the models do not separate as clearly as they did in the [O I]–FLR diagram (Figure 3, right panel). However, the H II models all lie below the demarcation line, and the NLR models mostly track the behavior of the [Ne V] sample. Running these models with different abundances did not significantly affect the results. The models shifted slightly to higher [Ne III]/[O II] ratios with lower metallicities, but the overall most influential factors were still the densities and ionization parameters used. We also tested this diagram by plotting the Seyfert galaxies found in Schmitt (1998; Figure 4), where all but two of the Seyfert galaxies are within the expected AGN upper right region.

Single-component models, like those discussed above, do not allow us to cover the full extent of our AGN sample.

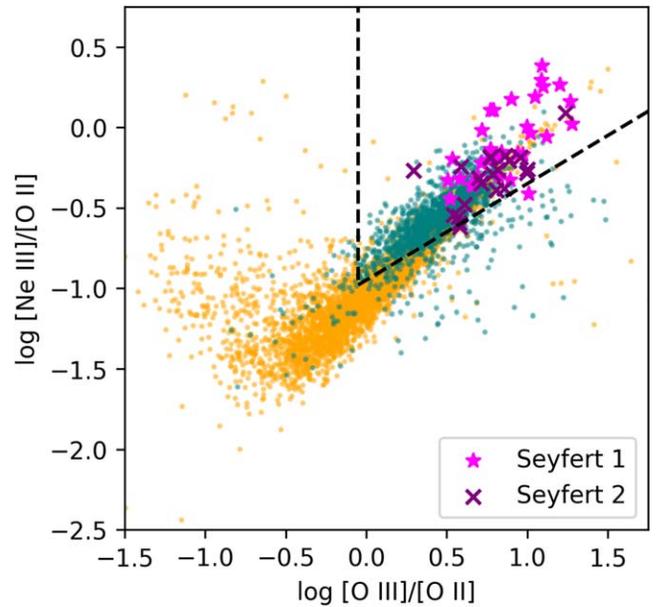

**Figure 4.** Diagnostic diagram from Figure 3 with the star-forming and [Ne V] samples, as well as the Schmitt (1998) Seyfert 1 and 2 AGN. All but two of the Seyfert galaxies are in the upper right region, where the AGN are expected to be found.

However, multicomponent models, such as the ones described in Kraemer et al. (1994), push the NLR results away from the demarcation line. Similar to the components in Kraemer et al. (1994), we adopt a three-component model with a range of high-to-low densities and ionization parameters. We use an INNER component with $\log U = -2$ and $\log n_H = 6$, a MIDDLE component with $\log U = -2$ and $\log n_H = 4$, and an OUTER component with $\log U = -3.5$ and $\log n_H = 3$. Varying the percentages of each component between 0% and 100% in increments of 10%, we create a grid of possible results using these three components, which are shown in Figure 5. We considered multiple combinations of densities, with none of the grids ever extending below the demarcation line. Model grids for SFGs on the [O II]–FLR diagram have been made by Jeong et al. (2020) using Cloudy and ionizing spectra from the Binary Population and Spectral Synthesis v2.2.1 models. Comparing these models to our data, their tracks cover the majority of our star-forming sample, with all of them below our demarcation line.

### 3.3. A 3D Diagnostic Diagram

The [O II]–FLR diagram contains the clear majority of the [Ne V] AGN within the AGN region. However, it still includes a large number of SFGs, due to there being significantly more SFGs than AGN in our sample, as well as the properties of the star-forming sample. Indeed, as we can see in Figure 2, the SFGs are biased toward the top part of the star-forming region, which indicates a bias toward galaxies with low metallicity. This contributes significantly toward pushing the SFGs into the AGN region of the [O III]–FLR diagram. Looking at the distribution of the [O III] luminosity for the SFGs and the [Ne V] AGN (Figure 6, top left panel), we can see that there is a clear distinction between the $L_{[\text{O III}]}$ ranges of the two samples. The overlap occurs mostly within $\log L_{[\text{O III}]} = 40$–42, which can also be clearly seen on the 3D plot (see Figure 6, bottom panel).

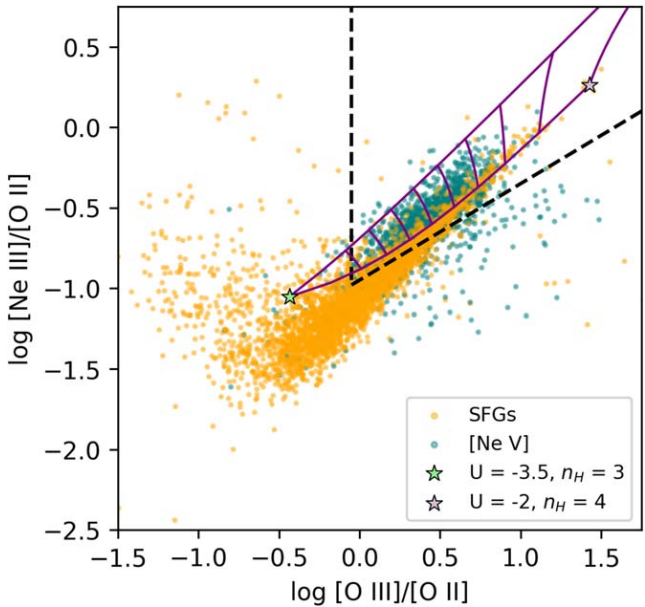

**Figure 5.** The [O II]–FLR diagram with the multicomponent model grid. The bottom line represents models with 0% of the INNER component, while the top line models have 0% of the MIDDLE component, and the rightmost vertical line uses 0% of the OUTER one. The percentage of the OUTER component used increases in increments of 10% for each consecutive vertical curved line going toward the left. The INNER and MIDDLE component percentages increase and decrease going up the vertical curved lines. Overall, the multicomponent model grid covers a wide area within the AGN region.

We want to use $L_{\text{[O III]}}$ to limit the number of SFGs in our AGN region. To find the optimal threshold, we calculate the number of galaxies from each sample left in the AGN region after different cutoffs (see Figure 6, top right panel). While the number of AGN stays almost constant all the way to log $L_{\text{[O III]}} = 41.5$, with a small decrease at the last cutoff, the number of SFGs drops significantly with each consecutive cutoff. Our recommendation to maximally limit the star-forming contamination percentage[11] of the AGN region is a log $L_{\text{[O III]}} = 42$ cutoff. This cutoff corresponds to a bolometric luminosity of 44.7 in log space when using the 454 factor from Lamastra et al. (2009), which corresponds roughly to the threshold between Seyfert galaxies and quasars. Using this cutoff decreases the contamination from 46% down to 10% while retaining 95% of the [Ne V] AGN in the AGN quadrant.

### 3.4. Limitations

This 3D [O II]–FLR diagnostic diagram (Figure 6, bottom panel) has some limitations. For instance, it will be more efficient when it comes to classifying Seyfert 1s compared to Seyfert 2s, as can be seen in the distribution of the Schmitt (1998) AGN in Figure 4. Looking at the Seyfert 1 galaxies specifically, all but one are in this region. This is expected, as we are sampling higher-density gas in the NLR of Seyfert 1 AGN compared to Seyfert 2s (Kraemer et al. 1998). The higher densities suppress the [O II] (see Table 2), which drives them to the upper right of our [O II]–FLR diagram. In Seyfert

1s, a view of the high-ionization nuclear emission-line region (Nagao et al. 2000) may also dilute the contribution from the star-forming regions in the galaxy. In addition, a sometimes substantial fraction of the inner NLR cannot be viewed in type 2s, lowering the [Ne III]/[O II] ratio used in our FLR diagrams, pushing them down (Kraemer et al. 2011).

The Portsmouth data that we are using are based on an automatic processing pipeline, rather than carefully measured values for each galaxy individually. It is expected in this case for us to get some amount of scatter, and despite it, our method is able to recover most of the AGN in the sample. The [O II]–FLR diagnostic diagram is able to classify galaxies up to a redshift of $z \leqslant 1.06$, which almost doubles the redshift range as compared to the standard diagrams, which rely on the H$\alpha$ line. Additionally, it only uses narrow and isolated emission lines that do not need to be deconvolved (unlike [N II] and H$\alpha$). This method is also very reliable at removing AGN from a sample in the case where a purely star-forming sample is needed.

Introducing the luminosity cutoff increased the reliability of the diagram; however, it does limit the usefulness of the method for lower-redshift galaxies. The AGN in our sample are generally more luminous than the typical AGN in the local Universe. Looking at the local AGN sample from Kraemer et al. (2011), we find that only 3 of the 51 AGN have a log $L_{\text{[O III]}}$ greater than 42. Although our diagram's efficacy is limited for lower redshift, the standard BPT may be used for local galaxies with high reliability, and the [O II]–FLR without a luminosity cutoff will still be useful.

### 3.5. The OHNO Diagram

As we have seen, it is possible to use a purely FLR-based diagram in order to separate AGN from SFGs. However, if we allow one of the axes to contain H$\beta$, we can still increase the redshift range as compared to a regular BPT and create a highly reliable diagram. This diagram, using [Ne III]/[O II] versus [O III]/H$\beta$, is the OHNO diagram (Zeimann et al. 2015; Backhaus et al. 2022). Backhaus et al. (2022) presented the OHNO diagram using spectra from 91 galaxies using Hubble Space Telescope Wide Field Camera 3 G102 and G141 grism spectroscopy. They find a demarcation line given by

$$\log_{10}\left(\frac{\text{[O III]}}{\text{H}\beta}\right) < \frac{0.35}{2.8 \log_{10}\left(\frac{\text{[Ne III]}}{\text{[O II]}}\right) - 0.8} + 0.64, \quad (5)$$

which is plotted in Figure 7. This demarcation line does not fit well with our sample, as it includes a large percentage of the SFGs in the AGN region.

This diagram allows us to achieve the same range of redshift as the [O II]–FLR diagram, with the [O III] line having the longest wavelength once again. The [O III]/H$\beta$ ratio is used in the regular [N II]-based BPT diagram and is thus helpful when separating SFGs from AGN. When plotted on a regular BPT, the SFGs are concentrated toward higher [O III]/H$\beta$ (see Figure 2). This indicates that, as a whole, our star-forming sample is weighted toward higher-ionization H II regions. This is also demonstrated on the OHNO diagram, where the star-forming sample lies at higher [O III]/H$\beta$ ratios compared to the Cloudy H II models. The NLR model grid uses the same components as described in Section 3.2 and covers a significant section of the AGN region (see Figure 7).

---
[11] The star-forming contamination percentage corresponds to the number of SFGs, divided by the total number of galaxies in the AGN region, multiplied by 100.

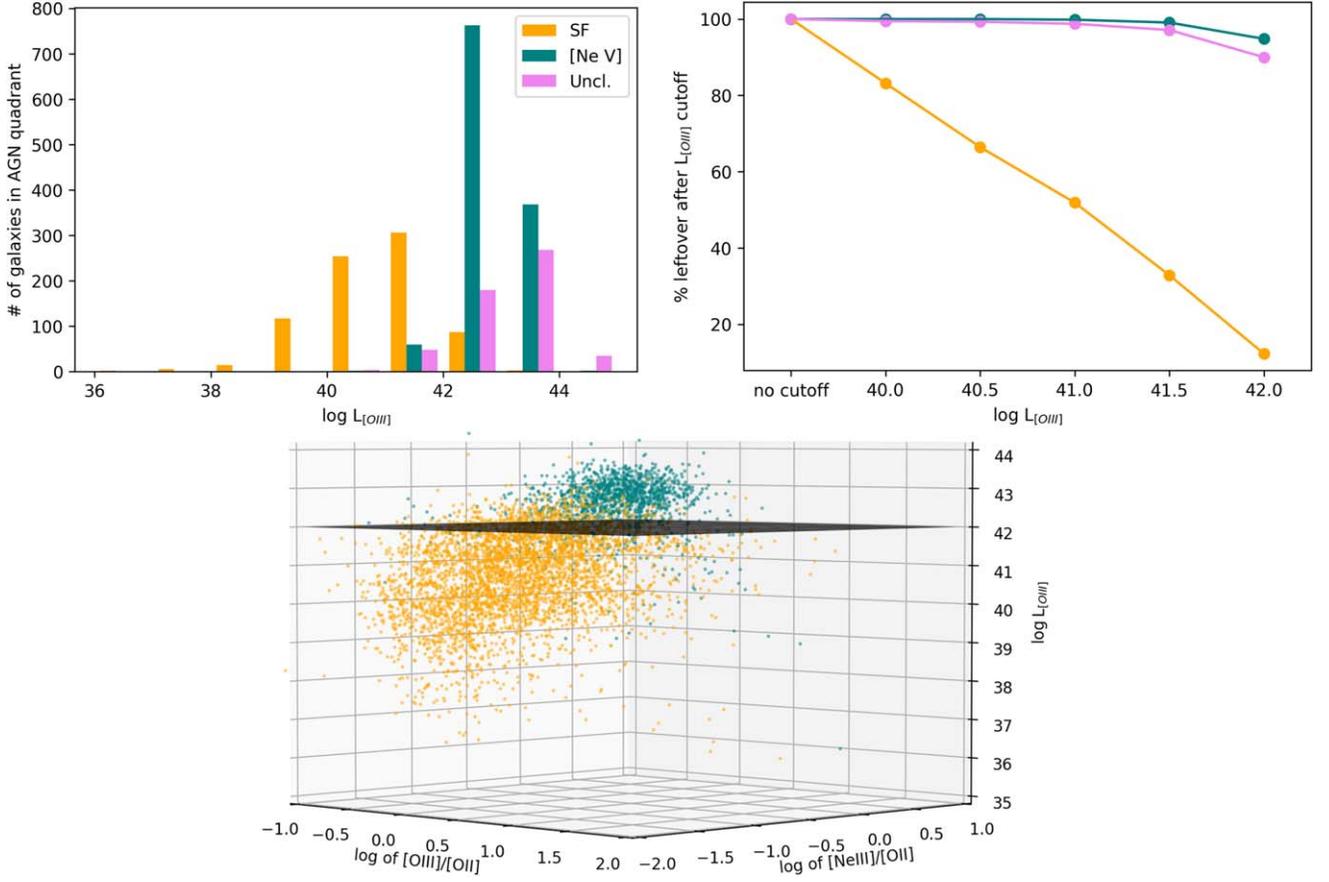

**Figure 6.** A demonstration of the effects of using an $L_{[O\,III]}$ cutoff in conjunction with the demarcation lines in order to further diminish the number of SFGs in the AGN region. Top left panel: distribution of the log of $L_{[O\,III]}$ for the star-forming, [Ne v], and unclassified samples. Top right panel: percentage of galaxies leftover in the AGN region of our [Ne III]/[O II] vs. [O III]/[O II] diagram after different log of $L_{[O\,III]}$ cutoffs. While the percentage of SFGs decreases steadily with increasing luminosity cutoffs, the number of [Ne v] and unclassified galaxies stays relatively constant. Bottom panel: a 3D diagram with the log of $L_{[O\,III]}$ on the z-axis, with a plane demonstrating the $L_{[O\,III]} = 42$ threshold. We can see once more that the [Ne v] AGN are at the higher end of the $L_{[O\,III]}$ axis, and that the two samples very clearly separate.

The demarcation line arising from our galaxy data is defined as

$$\log_{10}\left(\frac{[\text{Ne III}]}{[\text{O II}]}\right) < 1.8\left(\log_{10}\left(\frac{[\text{O III}]}{\text{H}\beta}\right) + 0.55\right)^2 - 3.4 \quad (6)$$

and shown in Figure 7. The AGN region of this plot contains 85% of the [Ne v] sample, while only including 2% of the SFGs. The high percentages are indicative of the reliability this method offers.

While this OHNO diagram is highly efficient at separating SFGs from AGN, it does have some limitations. Contrary to the [O II]–FLR diagram, the OHNO diagram may not work as well when it comes to type 1 Seyfert galaxies due to the need to deconvolve the broad and narrow components of H$\beta$. This method also does not allow us to find special kinds of galaxies, as will be explained in Section 4.3.

## 4. Applications

### 4.1. The Rest of the AGN Sample

The use of [Ne v] in our AGN selection process leads to an unambiguous AGN sample. However, it may lead to some biases, such as preferentially selecting high-luminosity AGN. We now explore the rest of the full AGN sample, those classified as AGN on the BPT diagram with [Ne v] S/N < 3.

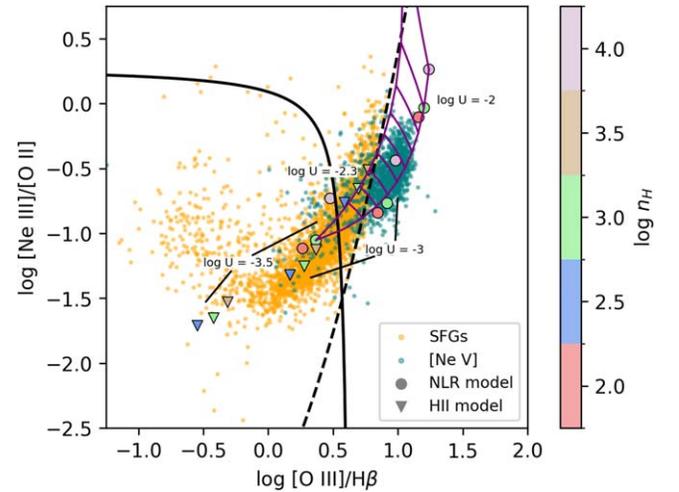

**Figure 7.** The OHNO diagram, with H II region and NLR single-component models, as well as the NLR multicomponent model grid. The solid black demarcation line comes from Backhaus et al. (2022) but does not work well to separate our data. The demarcation line used to isolate our [Ne v] AGN from the SFGs is the dashed line and is given in Equation (6).

We plot these AGN on the four different diagrams and separate the galaxies between Seyfert, LINER, and composite galaxies in Figure 8. We can easily see that in all four cases, the diagrams work well for the Seyfert galaxies while excluding the majority of LINERs and composite galaxies.

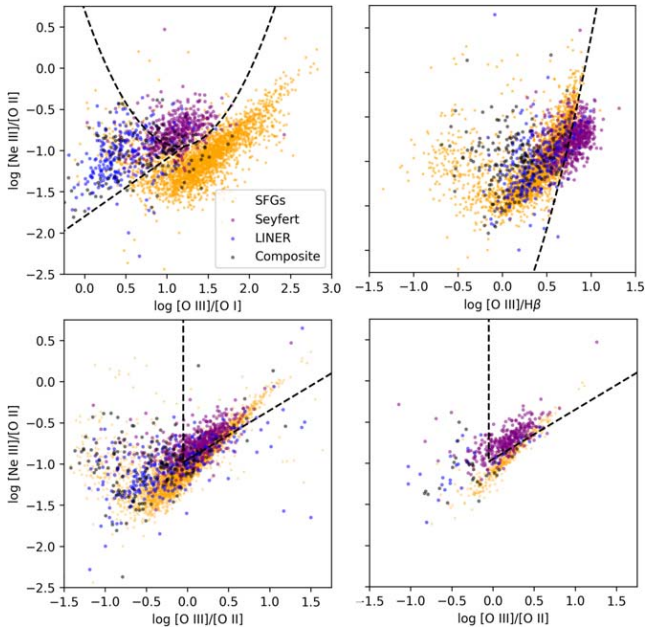

**Figure 8.** The AGN sample with [Ne v] S/N < 3. The AGN are separated by type, and we can see that overall, the Seyfert galaxies are consistently in the AGN region. The upper left plot corresponds to the [O I]–FLR diagram, while the upper right is the OHNO diagram. The bottom plots are the [O II]–FLR and 3D [O II]–FLR diagrams from left to right.

We have additionally summarized the percentages of AGN contained within the AGN region of each plot based on their BPT type (see Table 3). Overall, we can see that all of these diagrams are very efficient at finding the Seyfert galaxies, whether they have strong evidence of [Ne v] or not. They are not as efficient for composite galaxies and LINERs with [Ne v] and fail at picking up those weaker ones without [Ne v].

The composite galaxies are to a large extent ending up in the same region as the SFGs. This could be due to high SFRs and a subsequently increased [O II]. It should be noted that the demarcation line on the original BPT diagram is not universally accepted. For instance, the Kewley et al. (2001) extreme starburst demarcation line (see Figure 2) would classify almost all of our composite galaxies as SFGs. We can also see both by eye and by looking at the table that LINERs are also missing from our AGN region. These galaxies are fairly weak AGN that may or may not have enhanced star formation in the host. As has been shown in Meléndez et al. (2008), weak AGN with star formation will more strongly exhibit characteristics of SFGs rather than AGN.

Another interesting feature of the full AGN plots is the separation of the LINERs from the Seyferts and SFGs in the [O I]–FLR diagram. While the Seyfert galaxies are in the expected AGN region, the LINERs fall to the left of our demarcation line, away from the majority of the SFG sample as well. We can thus define a LINER region on the [O I]–FLR diagram. This additional demarcation line is shown in the top left panel of Figure 8 and given by

$$\log_{10}\left(\frac{[\text{Ne III}]}{[\text{O II}]}\right) > 0.7 * \log_{10}\frac{[\text{O III}]}{[\text{O I}]} - 1.80, \quad (7)$$

which is valid for $\log_{10}([\text{O III}]/[\text{O I}]) < 1.25$. This region, below the parabola but above the line given by Equation (7),

**Table 3**
Summary of Percentages of Each Type of Galaxy Present in the AGN Region of the Three Diagnostic Diagrams

| | | Percentage in the AGN Region | | | |
|---|---|---|---|---|---|
| Sample | BPT | [O I] FLR | [O II] FLR | 3D [O II] FLR | OHNO |
| SF | SF | 2 | 20 | 20 | 2 |
| [Ne v] AGN | Full | 89 | 83 | 91 | 83 |
| | Seyfert | 95 | 92 | 93 | 89 |
| | LINER | 55 | 18 | 33 | 40 |
| | Composite | 32 | 14 | 33 | 9 |
| Non-[Ne v] AGN | Full | 44 | 40 | 40 | 38 |
| | Seyfert | 68 | 61 | 74 | 63 |
| | LINER | 8 | 12 | 7 | 7 |
| | Composite | 19 | 9 | 9 | 1 |

**Note.** The number of SFGs between the [O II]–FLR diagram and the 3D [O II]–FLR diagram decreases significantly due to the [O III] luminosity cutoff (see Figure 8, bottom panels).

contains 85% of the LINERs, 62% of the composite galaxies, another 25% of the Seyfert galaxies, and only 9% of the star-forming sample.

### 4.2. The Unclassified Sample

The usefulness of the 3D [O II]–FLR and OHNO diagrams is apparent when considering our unclassified galaxies, for which Hα is shifted out of the bounds of the SDSS spectra. These galaxies all have a high redshift, with $z > 0.57$, and as a higher-$z$ sample will be biased to more luminous galaxies, the unclassified galaxy sample is likely to contain AGN. As we see in the top right panel of Figure 6, increasing the $L_{[\text{O III}]}$ does not affect the number of unclassified galaxies much, just like the [Ne v] sample.

Going back to the full unclassified sample and applying the 3D [O II]–FLR diagram limits to it, we get 485 AGN to add to our sample. Of these 485 galaxies, 293 of them have S/N > 3 for [Ne v]. Looking at the OHNO diagram, 526 galaxies are in the AGN region, with 318 galaxies also containing [Ne v]. There are also 122 galaxies with [Ne v] that are not in either region. By visual inspection of the 122 galaxies that are outside of both AGN regions, there are a handful of type 1 Seyferts, and the [Ne V] detection is suspect for a majority of them. Combining all of these methods, we are able to add 822 AGN to our sample. Figure 9 provides a breakdown of the different additions to the AGN sample.

### 4.3. The SFGs in the AGN Region

When using our 3D [O II]–FLR diagram with the log $L_{[\text{O III}]} = 42$ cutoff, we are left with only 114 SFGs in the AGN region. Looking at individual spectra, we confirmed that the [O III] emission lines were prominent and that the galaxies did not show any particular signs of being AGN (no [Ne v] or He II lines). Looking at the images for these galaxies in SDSS, a large number of them appear as either green or purple compact galaxies, each of which fits the description for green pea (Cardamone et al. 2009) and purple grape galaxies (Liu et al. 2022). Green peas are characterized by their low metallicity, intense star formation, and strong [O III] emission and are typically found in the redshift range $0.11 < z < 0.36$. The purple grapes have similar features but are found at different redshift ranges, namely, just below ($0.05 < z < 0.10$) and just above ($0.38 < z < 0.50$) the green peas.

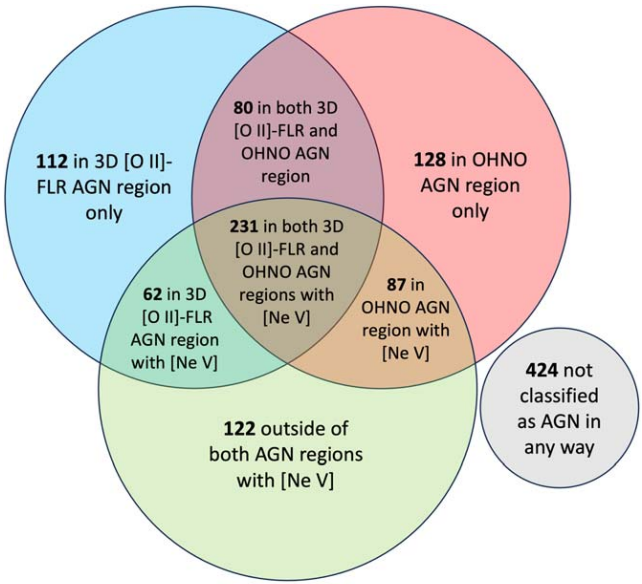

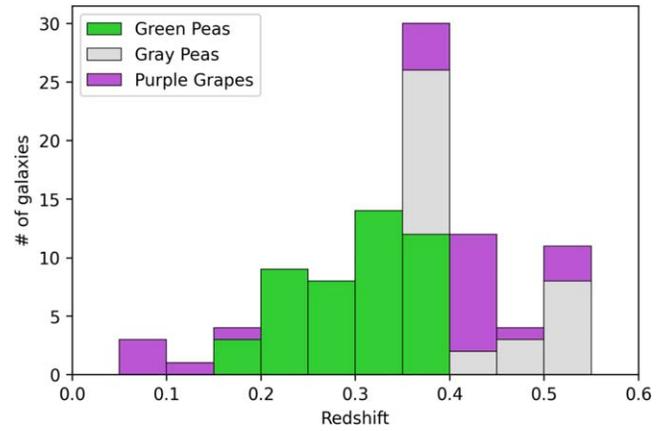

Figure 9. We now have three methods of classifying the unclassified galaxy sample: the [O II]–FLR diagram, the OHNO diagram, and the presence of the [Ne v] emission line. The Venn diagram outlines how many unclassified galaxies are classified as AGN using each of those three methods. The overlapping regions correspond to galaxies classified as AGN by more than one method. The bottom left panel is our 3D [O II]–FLR diagram, and the bottom right panel is the OHNO diagram, both with all unclassified galaxies in the sample plotted. Those with [Ne v] are marked with a black cross on top. As expected, the majority of the galaxies with [Ne v] fall within the AGN region of each diagram.

Figure 10. Redshift distribution of the SFGs in the AGN region of the 3D [O II]–FLR diagram. The color corresponds to the perceived color on the false-color SDSS images.

These galaxies exhibit similar features and spectra. The redshift span for our galaxies is shown in Figure 10 and agrees with the standard ranges. The differences in their coloring only have to do with their redshift, which pushes the most prominent emission lines ([O II], [O III], and H$\alpha$) in and out of the $g$, $r$, and $i$ SDSS filters. These filters are respectively mapped to blue, green, and red in SDSS false-color images. The green peas have the [O III] line firmly within the $r$ filter, with the other two contributing less to the color, leading to the green coloring. The purple grapes are found at two redshift ranges; the lower range leads to the [O III] line being in the $g$ filter and the H$\alpha$ in the $i$ filter. The second has the [O II] in the $g$ filter and the [O III] in the $r$ filter, with both cases leading to blue and red contributions, resulting in a purple image. The gray peas occur when the [O III] emission line falls between the $i$ and $z$ filters or the contributions from the green and red filters are similar.

When initially looking at these galaxies' spectra, we noticed signs of lower metallicities, e.g., weak [N II], as well as some indications of 4650 Å Wolf–Rayet (W-R) emission features

(Osterbrock & Cohen 1982). We also compared these spectra with those of known W-R galaxies and confirmed that they had very similar features. The [O III]/[O II] ratio used in our diagnostic has been investigated with regard to green pea galaxies by Jaskot & Oey (2013). By using Cloudy and Starburst99 models, they found that the only way to obtain high enough [O III]/[O II] ratios and the observed He II emissions was the presence of a substantial W-R population. This explains the striking similarity between the green peas' spectra and the known W-R galaxies. However, it should be noted that in general, local W-R galaxies do not exhibit [O III] luminosities as high as those seen in green peas (Brinchmann et al. 2008).

In Leitherer et al. (2018), the authors examine the super star cluster SSC-N and its surrounding H II region in the nearby dwarf galaxy II Zw 40. They find that the emission line ratios are most similar to green pea galaxies. We compare the H$\alpha$ luminosity for this star cluster ($9.2 \times 10^{40}$) to the average one for our green peas ($1.5 \times 10^{42}$) to estimate the number of W-R stars in our green peas. We find that, based on the difference in luminosity, we have on average about 15 times as many W-R stars. This suggests that the star-forming regions in the green peas are about 15 times bigger than the H II regions in dwarf galaxies of the sort described by Leitherer et al. (2018) and confirms that they generally have higher luminosity than their local analogs.

These special galaxies will thus be found in the region otherwise populated by AGN for several reasons. First, due to their high [O III] luminosities, they fall above the $L_{\text{[O III]}} = 42$ cutoff on the 3D [O III]–FLR diagram. Second, green peas have been found in the past based on their high [O III]/[O II] ratios (Jaskot et al. 2017), which also drives them into the AGN region. Finally, although the ones examined in this paper did not show signs of an AGN being present, past studies have demonstrated that green peas can sometimes also contain AGN (e.g., Harish et al. 2023). The green peas being in the AGN quadrant would then be expected in this case.

## 5. Conclusion

By applying several criteria to the full emissionLinesPort galaxy sample from the Portsmouth group, we separated it into a star-forming, an AGN, a [Ne v] AGN, and an unclassified sample. Using these four samples, we examine the use of a new diagnostic diagram to classify high-redshift galaxies.

1. We explored the relationship between the flux of [Ne v] and that of [Ne iii], [O iii], [O ii], and [O i]. We found that the correlation coefficients were high for all four lines. This confirmed our ability to use these lines in our FLR diagrams.
2. By plotting the [Ne iii]/[O ii] ratio against [O iii]/[O i], our [O i]–FLR diagram gives a different way of distinguishing between SFGs and AGN without the use of the H$\alpha$ emission line. However, as the [O i] line still has a longer wavelength, we are still faced with a limiting redshift of $z = 0.67$. Therefore, we also investigate the [O ii]–FLR diagram, which can be used for galaxies at high redshift (up to $z \leqslant 1.06$).
3. We used photoionization modeling to observe the physical basis for our FLR diagrams. We found that the models for the H ii region and the NLR separated well for both the [O i]–FLR and [O ii]–FLR diagrams. Multi-component models may also be used to cover a larger area of the AGN region, as they more accurately represent the physical conditions in the NLRs of AGN. The high-density component also allowed for the models to reach the Seyfert 1 galaxies from Schmitt (1998).
4. We added a third dimension to the [O ii]–FLR diagram in the form of a log $L_{[O III]} = 42$ luminosity cutoff. This allowed us to lower the star-forming contamination of the AGN region down to 10% while only losing 5% of the [Ne v] sample.
5. We investigated the OHNO diagram, which uses [Ne iii]/[O ii] against [O iii]/H$\beta$ (Backhaus et al. 2022). This diagram separated the star-forming sample and the [Ne v] galaxies very efficiently, and we suggested an updated demarcation line. However, its use may be limited in the case of type 1 Seyfert galaxies due to the narrow and broad components of H$\beta$ needing to be separated.
6. Applying the [O ii]–FLR diagram, the OHNO diagram, and the [Ne v] selection method allowed us to classify the previously unclassified galaxies and add 822 AGN to our sample.
7. Finally, by analyzing the SFGs left in the AGN region, we found that they are mostly green pea and purple grape galaxies. These galaxies are characterized by their high [O iii] luminosities, and we find that, based on their luminosities, they must contain 15 times more W-R stars than local dwarfs like II Zw 40.

The aim of this work was to search for alternative methods to the regular BPT diagram to classify galaxies at higher redshift. We were able to find two reliable ways to classify galaxies that both work up to a redshift of $z \leqslant 1.07$, as opposed to $z \leqslant 0.57$ for the standard BPT diagram. Having a more complete view of the number of AGN in the Universe is important for studying galaxy evolution and to allow large samples of AGN to be studied at higher redshift.

Another result of this analysis is the exploration of [Ne v] as an AGN tracer. Although this is well known for the local Universe (Abel & Satyapal 2008), reliable detection of [Ne v] can be used for AGN selection over a wide range of redshift. For example, [Ne v] can be detected with an S/N above 3 up to $z \leqslant 0.17$, 0.57, and 1 for [Ne v] luminosities of 40.22, 41.49, and 41.80 in log space.

### Acknowledgments

We would like to thank our referee, who gave us extremely insightful comments that led to great improvements to this paper.

This work has made use of SDSS DR12 data. Funding for the Sloan Digital Sky Survey IV has been provided by the Alfred P. Sloan Foundation, the U.S. Department of Energy Office of Science, and the Participating Institutions. SDSS-IV acknowledges support and resources from the Center for High Performance Computing at the University of Utah. The SDSS website is www.sdss4.org. The authors would like to thank the Portsmouth Group for making their data tables available online. The data can be found at https://www.sdss4.org/dr12/spectro/galaxy_portsmouth/.

This paper used the CLOUDY photoionization source code and its documentation, which may both be found at https://gitlab.nublado.org/cloudy/cloudy/-/wikis/home. We have also made use of the NASA/IPAC Extragalactic Database (NED), which is operated by the Jet Propulsion Laboratory, California Institute of Technology, under contract with the National Aeronautics and Space Administration.

### Appendix
### Full AGN Sample

We provide a table containing the full [Ne V]-selected AGN sample, as well as the previously unclassified galaxies determined to be AGN in this paper. We primarily include information regarding the emission-line flux values, as well as the method used to classify them as AGN (see Table 4).

Table 4
Final AGN Sample List Containing the Original [Ne v]-selected Sample and the Previously Unclassified Galaxies Determined to Be AGN

| BPT Type | $z$ | R.A. (J2000) | Decl. (J2000) | [Ne v] flux | [Ne III] flux | [O I] flux | [O II] flux | [O III] flux | H$\beta$ flux | OHNO | [O II]–FLR | [Ne v] AGN |
|---|---|---|---|---|---|---|---|---|---|---|---|---|
| | | | | \multicolumn{6}{c}{($10^{-17}$ erg s$^{-1}$ cm$^{-2}$ Å$^{-1}$)} | | | |
| LINER | 0.469 | 7.813 | 0.040 | 1.432 | 0.936 | 2.805 | 11.603 | 20.917 | 8.286 | ⋯ | ⋯ | ⋯ |
| Seyfert | 0.419 | 172.440 | 61.836 | 16.637 | 17.209 | 39.865 | 30.971 | 234.740 | 38.806 | ⋯ | ⋯ | ⋯ |
| Composite | 0.265 | 20.325 | 6.249 | 1220.885 | 160.503 | 8.011 | 518.546 | 80.368 | 79.184 | ⋯ | ⋯ | ⋯ |
| Unclassified | 0.602 | 346.768 | 12.938 | 13.464 | 21.754 | 11.722 | 24.318 | 191.571 | 17.349 | Yes | Yes | Yes |
| Unclassified | 0.726 | 154.212 | 51.202 | 256.492 | 82.357 | 0 | 5.644 | 681.616 | 192.868 | No | No | Yes |

**Note.** The columns indicated by emission lines correspond to the fluxes corrected for extinction. Those with a valid BPT type were corrected using the H$\alpha$/H$\beta$ ratio, while H$\beta$/H$\gamma$ was used for the previously unclassified ones. The last three columns indicate which of the three classification methods are used to classify the unclassified galaxies as AGN.

(This table is available in its entirety in machine-readable form.)

**ORCID iDs**

Léa M. Feuillet https://orcid.org/0000-0002-5718-2402
Marcio Meléndez https://orcid.org/0000-0001-8485-0325
Steve Kraemer https://orcid.org/0000-0003-4073-8977
Henrique R. Schmitt https://orcid.org/0000-0003-2450-3246
Travis C. Fischer https://orcid.org/0000-0002-3365-8875
James N. Reeves https://orcid.org/0000-0003-3221-6765